\documentclass[preprint,prd,aps,tightenlines]{revtex4}

\usepackage{verbatim}
\usepackage{graphicx}
\usepackage[usenames]{color}
\usepackage{psfrag}
\usepackage[ps2pdf,pdfpagemode=None,pdfstartview={FitH}]{hyperref}
\usepackage{hyperref}

\def\beq{\begin{equation}}
\def\eeq{\end{equation}}
\def\bea{\setlength\arraycolsep{1.4pt}\begin{eqnarray}}
\def\eea{\end{eqnarray}}

\begin{document}


\title{How Many Universes Do There Need To Be?}

\author{Douglas Scott} \email{dscott@phas.ubc.ca}
\author{J. P. Zibin} \email{zibin@phas.ubc.ca}
\affiliation{Department of Physics \& Astronomy\\
University of British Columbia,
Vancouver, BC, V6T 1Z1  Canada\bigskip\bigskip}

\begin{abstract}
\bigskip
\bigskip
\hrule
\smallskip
\smallskip
{\bf \noindent
In the simplest cosmological models consistent with General Relativity,
the total volume of the Universe is either finite or infinite, depending
on whether or not the spatial curvature is positive.  Current data
suggest that the curvature is very close to flat, implying that one can
place a lower limit on the total volume.  In a Universe of finite age,
the `particle horizon' defines the patch of the Universe which is observable
to us.  Based on today's best-fit cosmological parameters it is possible to
constrain the number of observable Universe sized patches,
$N_{\rm U}$.   Specifically, using the new {\sl WMAP\/} data, we can say that 
there are at least 21 patches out there the same volume as ours, at 
95\% confidence.  Moreover, even if the precision of our cosmological
measurements continues to increase, density perturbations at the particle
horizon size limit us to never knowing that there are more than
about 10$^5$ patches out there.}
\smallskip
\smallskip
\hrule
\end{abstract}

\maketitle


Understanding how big all of space is and our place within its immensity,
has long been one of the great
mysteries for human beings to ponder.
Within simple cosmological models, consistent with General Relativity, the
total volume depends on the curvature of the spatial sections, while the
volume of the Universe which is accessible to observation is determined by
the expansion history of the Universe.  Since we now have good empirical
knowledge of both global curvature and cosmological dynamics,
we can constrain the fraction of the Universe comprised by our observable
`patch'.

Although introductory cosmology courses teach the principles of non-Euclidean
spaces, we now know that our Universe is very close to having flat spatial
geometry.  The newest results from the Wilkinson Microwave Anisotropy Probe
({\sl WMAP\/}, \cite{wmap3}) measurements of the cosmic microwave background 
(CMB) combined with other cosmological data show that the average density of 
the Universe is within a few percent of the value required for flatness.

Sometimes it is assumed that the Universe {\em is\/} spatially flat.
However, this is an empirical question, and right now we do not know whether
the Universe is slightly closed or slightly open.  A closed universe, 
appealing on theoretical grounds (see \cite{closed} and
references therein), would have finite volume, while an open or
exactly flat space would have infinite volume.  Because of this, we do not 
know whether our own observable part is a
negligible or significant fraction of the whole volume, although as we discuss
below, it is now possible to place a rather robust limit on this fraction.

Tegmark has used the claim that the Universe is known to have infinite 
volume to argue \cite{Tegmark03} that there are as many `observable universe'
patches as there are `multiverses' in the quantum mechanical
`many-worlds' picture, and hence we
are already forced to imagine $\gtrsim 10^{100}$ copies of `the Universe',
perhaps more than the number of different possible particle configurations.  
Hence the suggestion is that modern cosmological observations lead one to
imagine that there are many near-copies of our patch out there in the vastness
of the entire Universe.  However,
as we will see below, the empirically-motivated limit is more like 10 than
$10^{100}$.  Hence there is no reason, purely on the grounds of cosmological
observations, to
be forced to believe in other versions of oneself, differing by only a few
bits of information.

   Of course, we have no possibility for observationally determining the 
character of the Universe on scales larger than the (apparent) particle 
horizon -- that is, we cannot see farther away (using photons at least)
than the last scattering 
surface, where the CMB radiation was 
released.  Nevertheless, we {\em can\/} determine that the observable volume 
is very close to homogeneous and isotropic, and, by applying the 
cosmological principle, conclude that the Universe continues to be well 
approximated by a homogeneous and isotropic Robertson-Walker (RW) 
metric on scales much larger still.  As we stated above, such a 
spacetime can be spatially open and infinite, in which case the arguments 
of Tegmark may be relevant, or it can be spatially closed and finite.  
In this latter case, our observable volume is a finite fraction 
of the total volume, and those arguments are invalid.

   If the universe is closed, current cosmological observations 
put an upper limit on the spatial curvature, and hence can be used to 
place a lower limit on the radius of curvature of such a model.  By 
using an observational probability distribution for curvature (together 
with other relevant cosmological parameters), and treating 
the Universe globally as an RW spacetime, it becomes a 
straightforward problem to determine the minimum ratio of the total 
volume to the observable volume consistent with observations, i.e. 
the minimum number of observable-universe-sized `patches' that must exist.

Based on growing circumstantial evidence, together with strong theoretical
motivation, there is good reason to believe that a period 
of inflation in the early Universe drove
the spatial curvature very close to zero.  
However, we take the approach here of determining what can be said about 
the curvature based on {\em empirical\/} evidence alone, coupled 
with the minimal extra ingredient of the assumption of global 
homogeneity and isotropy.  Indeed we will see that inflation itself 
{\em predicts\/} that we will never be able to determine the spatial 
curvature precisely.  In addition, non-trivial topologies can render a 
spatially flat universe finite in volume, but that possibility 
only strengthens our contention that we cannot conclude that the 
Universe is spatially infinite.  Current limits on the topology size are
in fact a little larger than the particle horizon distance \cite{small}.

   The metric for a spatially closed RW cosmology can be written
\beq
ds^2 = a^2(\eta)\left(-d\eta^2 + d\chi^2 + \sin^2\chi\,d\Omega^2\right),
\label{metric}
\eeq
where $\eta$ is the conformal time, $d\Omega^2$ is the solid angle element, 
$\chi$ is the comoving radial coordinate which runs from zero at our 
location to $\pi$ at the opposite `pole' of a homogeneous closed spatial 
slice, and we have set $c = 1$.
In this particular form of the metric, the scale factor
$a$ is not arbitrary, but in fact equals the physical radius of the spatial 
slices.  This conformal form makes it trivial to calculate 
the comoving distance to the particle horizon, i.e. the greatest value 
$\chi_{\rm ph}$ within our past light cone.  The result is
\beq
\chi_{\rm ph} = \int_0^{\eta_0}d\eta = \int_0^{a_0}\frac{da}{\dot{a}a} 
 = (-\Omega_{\rm K})^{1/2} D(\Omega_{\rm M}, \Omega_{\gamma}, \Omega_\Lambda),
\label{parthor}
\eeq
where the subscript ${}_0$ indicates a current value.  Here 
$\Omega_{\rm M}$, $\Omega_{\gamma}$, and $\Omega_\Lambda$ are the 
present-day energy densities in matter, radiation, and cosmological 
constant, respectively, as fractions of the critical density today, 
$\rho_{\rm crit} \equiv 3H_0^2/8\pi G$.  The parameter $\Omega_{\rm K}$ 
measures the contribution of curvature to the dynamics ($\Omega_{\rm K} 
= 1 - \Omega_{\rm M} - \Omega_{\gamma} - \Omega_\Lambda$ by the energy 
constraint equation).  The integral $D$ is given by
\beq
D(\Omega_{\rm M}, \Omega_{\gamma}, \Omega_\Lambda)
   = \int_0^1 \left(\Omega_{\gamma} + \Omega_{\rm M} a + \Omega_{\rm K}a^2
   + \Omega_\Lambda a^4\right)^{-1/2} da.
\label{ddef}
\eeq

   The volume of the current spatial slice out to distance 
$\chi_{\rm ph}$ is
\beq
V_{\rm ph} = \int\sqrt{h}d^3x = \pi a_0^3\left[2\chi_{\rm ph} - 
\sin(2\chi_{\rm ph})\right],
\eeq
where $h$ is the determinant of the spatial part of the metric 
(\ref{metric}).  Thus in particular the volume of the entire slice 
is $V_{\rm tot} \equiv 2\pi^2a_0^3$, and so the number of particle horizon 
volumes that can fit in the entire slice is
\beq
N_{\rm U} \equiv \frac{V_{\rm tot}}{V_{\rm ph}}
    = \frac{2\pi}{2\chi_{\rm ph} - \sin(2\chi_{\rm ph})}.
\label{nudef}
\eeq

   The parameters $\Omega_{\rm M}$, $\Omega_{\gamma}$, and 
$\Omega_\Lambda$ are now well-measured \cite{wmap3} and they fix 
$\Omega_{\rm K}$ through the energy constraint equation, assuming 
Einstein gravity.  Choosing the best-fit WMAP 3 year data set 
parameters $\Omega_{\rm M} = 0.24$, $\Omega_{\gamma} = 5.5\times10^{-5}$, and 
$\Omega_\Lambda = 0.76$, we find, performing the integral (\ref{ddef}), 
that $D \simeq 3.5$.  This tells us that the 
apparent particle horizon exceeds the current Hubble radius by 
the factor $3.5$.  Note that this calculation, through Eq.~(\ref{ddef}), 
depends on the assumption of radiation domination into the arbitrary 
past.  Of course standard inflationary models violate this assumption, 
and the particle horizon can actually diverge in these models.  But 
we are interested here in the {\em apparent\/} particle horizon,
i.e.~the comoving distance to the last scattering surface, as this determines 
the size of the {\em observable\/} Universe.

\begin{figure}[pb]\begin{center}
\includegraphics[width=8cm]{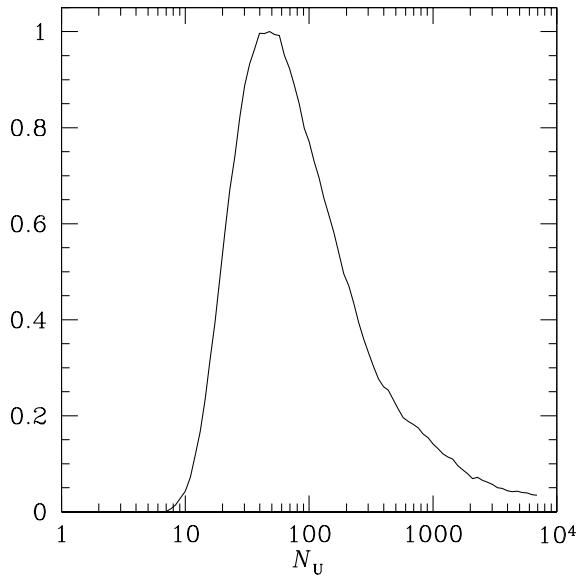}
\caption{Probability distribution (in arbitrary units) for the number of 
universe patches $N_{\rm U}$ based on {\sl WMAP} + {\sl HST} data.  The 
$95$\% confidence lower limit is $N_{\rm U} > 21$.  Although by eye this 
distribution appears to favour values near $N_{\rm U} \sim 100$, this is 
actually not the case due to the logarithmic scale and because 
of a large peak at infinite $N_{\rm U}$ corresponding to open models.}
\label{limitfig}
\end{center}\end{figure}

   We can estimate the limit on the number of observable Universe patches 
by looking at the observed probability distributions for the cosmological 
parameters, and through Eqs.~(\ref{parthor}) and 
(\ref{nudef}), convert these to a distribution for $N_{\rm U}$.  Since, 
of the relevant parameters, $\Omega_{\rm K}$ has the greatest relative 
uncertainty, the uncertainty in $N_{\rm U}$ will be dominated by that 
of $\Omega_{\rm K}$ through Eq.~(\ref{parthor}).  Using the Markov 
Chain Monte Carlos provided by the {\sl WMAP\/} team for non-flat cosmologies
(together with additional information on the Hubble constant from the Hubble 
Space Telescope ({\sl HST}) Key Project \cite{hst}) we 
calculated the likelihood function for $N_{\rm U}$; the result is presented 
in Fig.~\ref{limitfig}.  Based on this distribution we find a $95$\% 
confidence lower limit of $N_{\rm U} > 21$.  That is, $95$\% of models 
consistent with the {\sl WMAP\/} and {\sl HST} data are closed models with 
more than $21$ observable-universe-sized patches, or are open models.  
Different choices of cosmological data, model spaces, and Bayesian priors
yield somewhat different distributions for $N_{\rm U}$, but all
reasonable choices yield a lower limit of $N_{\rm U} \gtrsim 10$.

   Ultimately we are limited by cosmic variance in our observable patch.  
Due to the spectrum of perturbations produced by standard inflationary 
models, the curvature perturbation on the scale of our patch today is of 
order $10^{-5}$ (see also for example \cite{knox}),
and hence we are unlikely to ever know with confidence that
$|\Omega_{\rm K}| < 10^{-4}$ (since we could live a few $\sigma$ into
the tail of the distribution).  In that case Eq.~(\ref{nudef}) becomes
\beq
N_{\rm U} \simeq \frac{3\pi}{2\chi_{\rm ph}^3} 
          \simeq 0.1(-\Omega_{\rm K})^{-3/2}.
\eeq
Hence the {\it best\/} lower limit that we will ever be able to place is
$N_{\rm U}\gtrsim 10^5$, unless of course improved measurements find that the
Universe is actually open.  But assuming that we continue to measure that
the curvature is close to flat, then cosmology alone will only enable
us to infer that there are {\em at least\/} 100{,}000 other observable
patches out there.  Determining that the Universe is genuinely infinite will
remain beyond the reach of purely empirical studies.

\section*{Acknowledgments}

We are grateful to Yuna Kotagiri, Max Tegmark, Tom Waterhouse, and Martin
White for useful discussions.  This work was supported by the Natural 
Sciences and Engineering Research Council of Canada.

\end{document}